\documentclass[journal]{IEEEtran}

\IEEEoverridecommandlockouts
\usepackage{cite}
\ifCLASSOPTIONcompsoc
\usepackage[caption=false,font=normalsize,labelfont=sf,textfont=sf]{subfig}
\else
\usepackage[caption=false,font=footnotesize]{subfig}
\fi
\usepackage{amsmath,amssymb,amsfonts}
\usepackage{nicefrac}
\usepackage{algorithmic}
\usepackage{optidef}
\usepackage{graphicx}
\usepackage{textcomp}
\usepackage{balance}
\usepackage{xcolor}
\usepackage{soul}
\usepackage{paralist}
\usepackage{booktabs}
\usepackage{textcomp}
\ifCLASSOPTIONcompsoc
 \usepackage[caption=false,font=normalsize,labelfont=sf,textfont=sf]{subfig}
\else
 \usepackage[caption=false,font=footnotesize]{subfig}
\fi

\def\BibTeX{{\rm B\kern-.05em{\sc i\kern-.025em b}\kern-.08em
    T\kern-.1667em\lower.7ex\hbox{E}\kern-.125emX}}

\graphicspath{{images/}}
\DeclareGraphicsExtensions{.jpg,.png,.png}
\usepackage[ruled,vlined]{algorithm2e}
\usepackage[version=4]{mhchem}
\usepackage{multirow}
\usepackage{siunitx}
\makeatletter
\def\hlinewd#1{
  \noalign{\ifnum0=`}\fi\hrule \@height #1 \futurelet
   \reserved@a\@xhline}
\makeatother

\begin{document}

\title{Designing a Transactive Electric Vehicle Agent with Customer's Participation Preference}

\author{\IEEEauthorblockN{Ankit Singhal,~\IEEEmembership{Member,~IEEE,}, Sarmad Hanif,~\IEEEmembership{Member,~IEEE,}, Bishnu Bhattarai,~\IEEEmembership{Senior Member,~IEEE,}, Fernando B. dos Reis,~\IEEEmembership{Member,~IEEE,}, Hayden Reeve,~\IEEEmembership{Member,~IEEE,}, and Robert Pratt,~\IEEEmembership{Senior Member,~IEEE}}
\thanks{This research is supported by PNNL with funding from the U.S. Department of Energy under contract No. DE-AC05-76RL01830.}}

\maketitle
\pagestyle{plain}
\begin{abstract}
The proliferation of electric vehicles (EVs) and their inherent flexibility in charging timings make them an asset to improve grid performance. In contrast to direct control by a utility or autonomous price-based charging, the transactive control framework not only provides benefits to both grid and customers but also ensures customer autonomy. In this work, we design a transactive electric vehicle (TEV) agent that incorporates the EV owner's willingness to trade-off between savings and amenity in form of a slider, where the EV owner's amenity is characterized as vehicle readiness. Further, a privacy-preserving bidding formulation is proposed that also represents the customer's transactive preference. A transactive market mechanism is discussed that integrates the TEV Agents into the local retail market and reconciles with the current day-ahead and real-time market structure. It is demonstrated that the proposed slider is able to provide a preferred trade-off between savings and amenity to individual customers. At the same time, the market mechanism is shown to successfully reduce both peak prices and peak demand. A comparative investigation of V1G and V2G technologies with respect to the battery prices is also discussed.

\end{abstract}


\IEEEpeerreviewmaketitle
\section{Introduction}
Decarbonization directives, aggressive EV adoption mandates, falling battery prices, improved driving ranges, and increasing availability of public charging stations are some of the factors fueling EV growth worldwide \cite{michael_woodward_electric_2020,noauthor_summary_2019}. The resulting increase in the electricity demand to charge these EVs may pose significant challenges to the power system in the form of transformer overloading, power congestion, increased peak load, and power losses \cite{shafiee_investigating_2013,clement-nyns_impact_2010,godina_smart_2016}.
These adverse impacts are usually mitigated by coordinated EV charging scheduling that takes advantage of the flexibility in vehicle charging timings. These methods can be broadly categorized into the following two categories: (1) direct control and (2) incentive-based control.

In direct control strategies, a central controller or an EV aggregator directly schedules the charging of EVs to improve grid performance based on prior financial agreements with the owners \cite{al-ogaili_review_2019}. A rich literature is available in this category, where the aggregator collects data from EV owners to determine a system-wide optimal schedule to provide various grid services \cite{clement-nyns_impact_2010,zhang_methodology_2012,hua_adaptive_2014,amamra_vehicle--grid_2019, faridimehr_stochastic_2019, mohamed_real-time_2014}. For instance, a quadratic optimization is proposed to determine an optimal EV scheduling to minimize grid losses and flatten load profile by \cite{clement-nyns_impact_2010} and \cite{zhang_methodology_2012}, respectively. Similarly, \cite{hua_adaptive_2014} proposes a DC power flow-based adaptive EV scheduling optimization to prevent voltage and thermal violations in the grid under high EV penetration. \cite{amamra_vehicle--grid_2019} proposes a centrally optimized EV scheduling to provide frequency and voltage regulation services to the grid. 
Although direct control methods are able to obtain optimal grid performance, the decision-making authority does not remain with the customers, which raises concerns of consumer autonomy and privacy protection.
Indeed, the requirement of customers' private information remains one of the major challenges of rolling out mass adoption in grid-friendly services \cite{dobelt_consumers_2015}.

In incentive-based control strategies, EV owners maximize their preferences (e.g., savings and comfort). These control methods are usually local in nature and make their decisions based on electricity price signals \cite{al-ogaili_review_2019,ma_decentralized_2013}. Some examples of such controls can be found in  \cite{wi_electric_2013, veldman_distribution_2015, al-ogaili_review_2019} where individual EV owners' cost is minimized and \cite{singhal_transactive_2021} where EV owners' comfort is also optimized along with cost. 
However, the customers' benefit in these control strategies may conflict with the grid's performance, as suggested in a comparative study by \cite{veldman_distribution_2015}. Further, the issue of congestion due to price-responsive EVs is well known \cite{remco2014}, where all EVs jump to charge during low price periods and end up causing peak load. Any intervention strategy by the utility to mitigate such issues will be unpopular among consumers because it challenges their autonomy to consume energy.

As the modern grid moves to decentralized and consumer-centric operations, consumer preferences and autonomy are important considerations for utilities in addition to safe grid operation.  The transactive energy (TE) framework has been shown great interest from the research community in this regard, as it combines the techniques from economics and controls to  coordinate flexible energy assets using value as a key operational parameter through transparent and competitive means \cite{jin_xin_challenges_2015, the_gridwise_architecture_council_gridwise_2015}. 
In the TE scheme, each distributed energy resource (DER) is controlled by its owner via a local transactive agent that leverages its flexibility and uses economic incentives to maximize its benefits. However, the transactive mechanism is designed to align individual customer's interests with the system needs \cite{DR_to_transactive_2017}, thus harnessing various grid support functionalities. As a result, it provides the advantages of both direct control and incentive-based control strategies. 
Recent progress on TE system development \cite{taft2016architectural, GWACRoadmap} and encouragement from directives such as FERC order 2222 are setting a foundation for the participation of small-scale flexible resources in retail electricity markets. To this end, the proposition of a transactive electric vehicle agent (TEV) to enable an EV to participate in the electricity market is the main objective of this work.

There is a body of work that explores the game-theoretic approaches to propose price-based distributed control to realize a transactive framework for EV scheduling, such as Nash equilibrium convergence \cite{ma_decentralized_2013, wu_vehicle--aggregator_2012, gan_optimal_2013}, Stackelberg game \cite{tushar_economics_2012}, etc. As much as these works prove the theoretical existence of an equilibrium and thereby a global optimal solution, the practical implementation of these ideas is extremely difficult \cite{lian_transactive_2018}. Moreover, these require knowledge of EV owners' private information that remains a challenge in real world. Some recent works \cite{liu_transactive_2019, galvan_transactive_2016, behboodi_electric_2016} propose transactive schemes where EVs participate in day-ahead (DA) or real-time (RT) markets to minimize charging cost, but with no proper consideration of EV owners' comfort and willingness. 

In summary, though the TE approach, in principle, promises to address the privacy and autonomy concerns, its application in EV integration literature has the following \textbf{major gaps}. First, all of the TEV work only attempts to minimize user cost without explicitly modeling customers' amenity (or comfort). Due to this, the EV owner's ability to choose the desired trade-off between cost savings and amenity is missing. In other words, the customer does not have an intuitive access to control its willingness to participate in the transactive market dynamically. Second, existing methods require private information sharing to an external entity (such as charging requirements, EV specifications), raising privacy protection issues. These gaps are not consistent with the vision of the transactive paradigm that ``the individual customer understand their needs best'' \cite{DR_to_transactive_2017}.




Therefore, we propose a TEV agent that models user preferences along with providing grid services by participating in both day-ahead (DA) and real-time (RT) retail electricity markets. Following are the \textbf{main contribution} of this work:
\begin{enumerate}
    \item We propose a novel TEV Agent that (a) explicitly characterizes vehicle readiness as a key amenity for an EV owner and (b) provides the customer an intuitive way to choose its willingness to participate in the transactive market, in form of a preference slider.
    \item We propose a novel demand bid formulation that is privacy-preserving, i.e., it masks the user preferences and their private information and merely shares a set of demand and price points to an external entity.
    \item We provide a mechanism for integration of TEV agent into a retail market, which is shown to reconcile with the current DA and RT structures of the wholesale market.
\end{enumerate}
 Further, Since vehicle-to-grid (V2G) technologies are considered in the future EV roll-outs, we demonstrate the applicability of the proposed method on V2G and investigate the comparative benefits with grid-to-vehicle (V1G) in a transactive market. 
 
The rest of the paper is organized as follows. Section~II presents an overview of the TEV agent framework. Next, Sections~III and IV present the modeling of EV behavior and customer preference, respectively. Section~V discusses the market mechanism to facilitate TEV bidding and the clearing process. Section~VI demonstrates the simulation case study and comparative analysis, followed by conclusions in Section~VII.

\section{Transactive EV Agent Framework}
EV owners need two-way communication with the distribution system operator (DSO) to participate in the transactive market. To this end, we propose a TEV agent that is a software agent associated with the EV owner and acts on behalf of the customer. It represents customer's interests in the transactive process, such as their preferences and flexibility. Needless to say, the EV owner will be required to enroll in the transactive program with the DSO, an entity that is assumed to be responsible for aggregation and retail market operation in this work.

\begin{figure}[t]
\centering
\includegraphics[width=1\columnwidth]{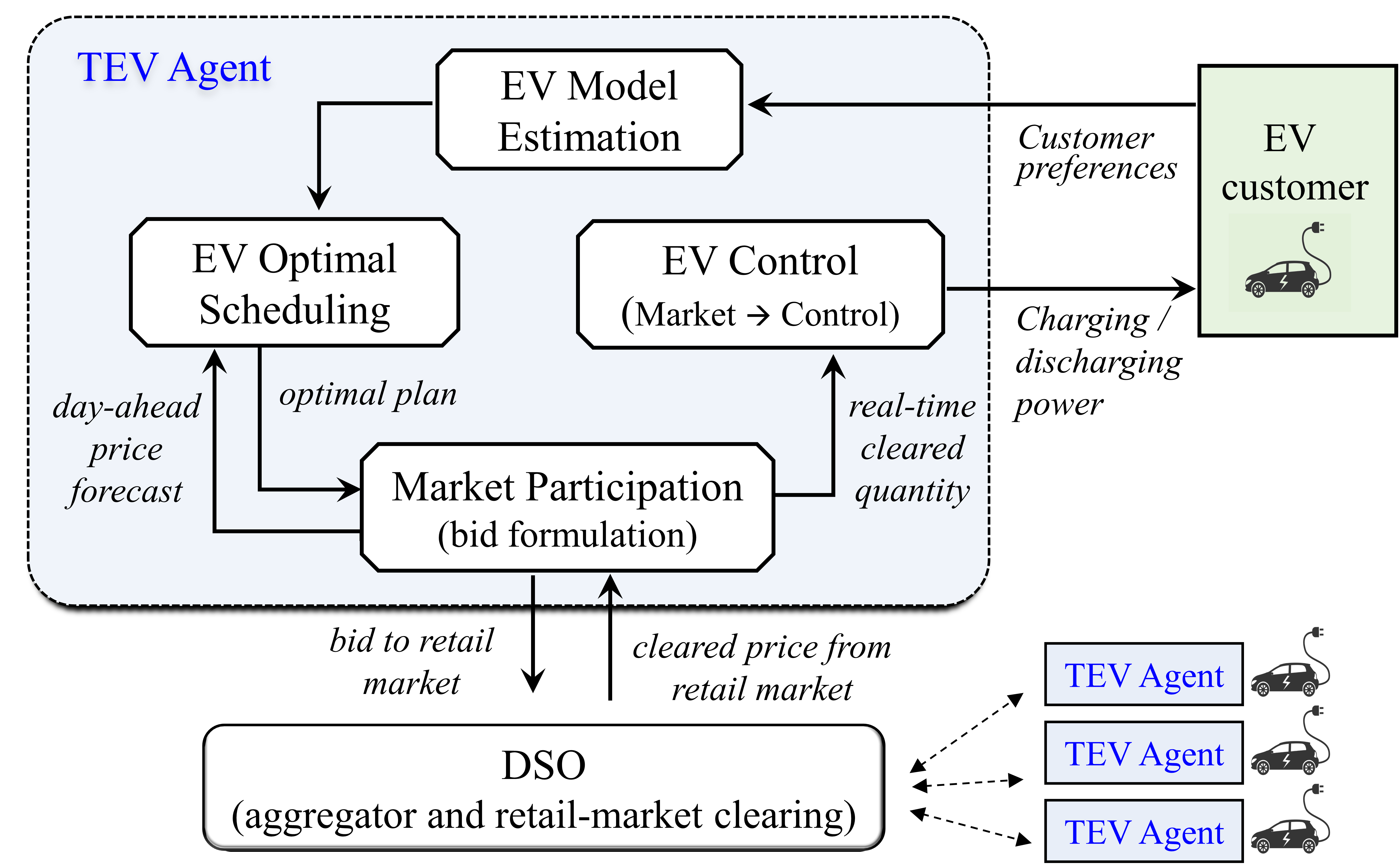}
\caption{A high-level schematic of the transactive EV agent with its functional modules and its interface with the rest of the system.}
\vspace{-3mm}
\label{Fig:EVSChematic}
\end{figure}

An overview of TEV agent is illustrated in Fig. \ref{Fig:EVSChematic}, where several TEV agents interface with DSO to participate in retail markets. To accomplish it's objectives, the TEV agent is designed with the following four modules:
\begin{inparaenum}
    \item EV model estimation,
    \item optimal EV scheduling,
    \item market bidding mechanism, and
    \item EV control.
\end{inparaenum}
The first module estimates the physical behavior and dynamics of the EV by monitoring state-of-charge (SOC) directly from the EV and taking user inputs such as daily driving schedule, EV specifications, and customer's preference level to participate in the transactive market. Note that the user inputs can be updated based on customer's need and these private information is not shared with any external agency. Then the EV scheduling module prepares an optimal plan (a charging schedule) the EV would ideally like to operate on to maximize its benefits based on the forecast electricity prices and the EV model estimation. 
The third module enables the TEV agent to participate in both DA and RT retail transactive market by interfacing with the DSO. It constructs a flexible price-quantity bid curve around the optimal schedule point as well as adjusts the optimal plan if forecast price differs from the retail cleared price. 
The last module of EV control takes the RT cleared quantity from the market module and sends the charging/discharging control signal to EV such that the EV operation meets the quantity committed to the market. Each of the modules and their processes are described in detail in the next sections.

\section{EV Model Estimation}
The EV model estimation requires primarily two sets of data i.e. physical parameters of the vehicle, and EV usage (driving) patterns as listed in Table \ref{tab:ev_parameters_list} and discussed below.

\begin{table}[]
\caption{Parameters list to estimate an EV model and their nomenclature}
\label{tab:ev_parameters_list}
\begin{tabular}{@{}p{2.8cm}lp{4.4cm}@{}}
\toprule
\multicolumn{3}{c}{Physical parameters} \\ \midrule
Range (miles) & $r$ & Maximum miles an EV can drive per full charge cycle \\
Mileage (miles/kWh) & $m$ & Discharge rate while driving \\
Charging rating (kW) & $E^{in}_{max}$ & Maximum energy a charger can transfer from grid to EV in an hour \\
Discharging rating (kW) & $E^{out}_{max}$ & Maximum power a charger can transfer from EV to grid in an hour \\
Charging efficiency & $\eta_{in}$ & Efficiency of charger while charging EV from grid in V1G mode \\
Discharging efficiency & $\eta_{out}$ & Efficiency of charger while discharging EV to grid in V2G mode \\
\toprule
\multicolumn{3}{c}{Vehicle usage pattern parameters} \\ \midrule
Plug-in time & $t_{in}$ & The latest time the car arrives home \\
Plug-out time & $t_{out}$ & The earliest time the car leaves home \\
Plug-in duration & $T_p$ & Time elapsed between plug-in and plug-out \\
Daily travel miles & $d$ & The miles responsible for SOC depletion daily \\ \bottomrule
\end{tabular}
\end{table}

\subsection{EV Physical Parameters}
The physical model of an EV can be sufficiently characterized by its range, charger ratings, and mileage as listed in Table \ref{tab:ev_parameters_list}. These physical parameters for EVs from different manufacturers are sourced from publicly available data \cite{ev_sale_2019, EV_data} and tabulated in Table \ref{tab:ev_model_sale_data}.  These are listed in decreasing order of their sales as a percentage of the U.S. total during 2016-2019. More than 80\% of EV drivers in the USA charge only at home due to convenience \cite{IDL2015}. Therefore, we assume only home charging in this work. Further, it is assumed that all residential EV owners have level 2 charging available as level 1 charging is too slow for new EV models with a longer range.

\begin{table}[h!]
\caption{Sale data for top 15 commercially available EVs during 2016-2019 and their corresponding physical parameters}
\label{tab:ev_model_sale_data}
\begin{tabular}{@{}lp{1cm}p{0.8cm}p{1.4cm}p{1cm}@{}}
\toprule
Electric vehicle model & Sale \% & Range (miles) & Charger rating (kW) & Mileage (miles/kWh) \\
\midrule
Tesla Model 3 & 44.11\% & 220 & 11.5 & 3.84 \\
Tesla Model S & 14.52\% & 285 & 11.5 & 3.33 \\
Tesla Model X & 12.92\% & 258 & 11.5 & 2.85 \\
Chevy Bolt & 8.66\% & 238 & 3.3 & 3.57 \\
Nissan Leaf & 7.79\% & 151 & 3.3 & 3.33 \\
BMW i3 & 3.70\% & 153 & 7.4 & 3.84 \\
VW e-Golf & 2.04\% & 125 & 7.2 & 3.57 \\
Fiat 500E & 1.48\% & 84 & 6.6 & 3.33 \\
Audi e-tron & 0.80\% & 204 & 11 & 2.17 \\
Kia Soul EV & 0.76\% & 111 & 6.6 & 3.22 \\
Ford Focus EV & 0.49\% & 115 & 6.6 & 3.22 \\
Smart ED & 0.46\% & 58 & 3.3 & 3.22 \\
Chevy Spark & 0.46\% & 84 & 3.3 & 3.57 \\
Jaguar I-Pace & 0.44\% & 234 & 7 & 2.27 \\
Honda Clarity BEV & 0.42\% & 89 & 7.7 & 3.33 \\
\bottomrule
\end{tabular}
\end{table}

\subsection{EV Usage Pattern}
In order to model EV charging and discharging behavior, the driving pattern of each EV is required. 2017 National Household Travel Survey (NHTS) data provide the residential car driving schedules in the form of their home arrival time ($t_{in}$), home leaving time ($t_{out}$) and daily travel miles ($d$) as listed in Table \ref{tab:ev_parameters_list}. These schedules are assumed to represent residential EV behavior where $t_{in}$ and $t_{out}$ are considered as plug-in and plug-out time, respectively.

In order to create a diverse EV population, a car model is picked at random from a distribution of EV models based from Table \ref{tab:ev_model_sale_data}. Then, a driving schedule from NHTS data is picked at random and attached to the selected car model while ensuring that (a) the daily miles traveled do not exceed the range of the EV model, i.e., $d<r$, and (b) the plug-in duration is sufficient to fully charge the EV before it leaves home every day, i.e., $T_p \times E_{max}^{in} > r/m $.


\subsection{Physical Equations and Constraints}
Let us define a set $T=\{1,2,...,N\}$ denoting a time horizon window of $N$ hours and a subset $T_{tran} \subset T$ that contains all transactive hours, i.e., when a vehicle is parked at home and available for charging or discharging. Further, subsets $T_{in} \subset T, T_{out} \subset T$ are defined as collections of the home arrival and departure hours, respectively.
$E^{in} (t)$ and $E^{out} (t)$ are the charging and discharging energy amount in $kWh$, respectively, exchanged by the EV with the grid. During non-transactive hours, the energy exchanged between grid and EV is zero: 
\begin{align}
    E^{in}(t)=E^{out} (t)=0 \;\;\; \forall t\in T-T_{trans} \label{eq:first_const}
\end{align}
During transactive hours, when EV is exchanging energy with the grid, the energy exchange can range between 0 and the maximum amount permissible by the charger rating:
\begin{align}
    0 \leq E^{in}(t) \leq E_{max}^{in} \;\;\; \forall t\in T_{trans} \\
 0\leq E^{out}(t) \leq E_{max}^{out} \;\;\; \forall t\in T_{trans} 
\end{align}
Note that $E^{out}_{max}$ is nonzero only in V2G case where the EV can discharge to provide energy to the grid. The base case charging strategy considers only V1G mode with constant charging rate ($E^{in}=E_{max}^{in}$) until the EV is fully (100\%) charged. The energy exchange recorded by the net meter is different from what the EV experiences, because it also accounts for losses. Metered energy, $E^{in}_b$ and $E^{out}_b$, responsible for billing in the case of charging and discharging respectively, can be obtained using their efficiency factors as:
\begin{align}
    E^{in}_{b} = E^{in} \div \eta_{in} \\
    E^{out}_{b} = E^{out} \times \eta_{out}
\end{align}
SOC of the vehicle at time $t$ is denoted by $C(t)$ and is governed by different equation during different time periods based on EV's operational state as shown below:
\begin{align}
    C(t)=C(t-1)-E_d/2  \;\;\; \forall t\in (T_{in} \cup T_{out}) \label{eq:driving_drain}
\end{align}
\vspace{-4mm}
\begin{equation}
    \begin{split}
     C(t)=C(t-1)+(E^{in}(t)-E^{out} (t))\\ \forall t\in T-(T_{in}\cup T_{out})
     \end{split} \label{eq:soc_gain}
\end{equation}
\vspace{-4mm}
\begin{align}
    C(t-1)=C^{max} \;\;\; \forall t\in T_{out} \label{eq:bound}
\end{align}
Equation (\ref{eq:driving_drain}) denotes the depletion of energy due to driving. For simplicity, without loss of generality, the daily driving energy depletion, $E_d=d/m$, is equally distributed at the departure ($t_{out}$) and arrival ($t_{in}$) hours. For all other hours, the change in vehicle SOC is denoted by (\ref{eq:soc_gain}). Additionally, at all hours, the SOC is bounded by it's maximum and minimum permissible values, $C^{max}$ and $C^{min}$, respectively as,
\begin{align}
    C^{min} \leq C(t) \leq C^{max} \label{eq:last_const}
\end{align}
\section{Optimal EV Scheduling}
\label{Sec:Methodology}
The optimal EV scheduling module provides an hourly optimal operational schedule for the EV charging for the horizon window of $N$ hours. The most crucial aspect of optimal scheduling of the proposed TEV agent is modeling the customer's preference level to participate in transactive market, as discussed in following subsections.

\subsection{Modeling Customers' Transactive Preference}
\begin{figure}
    \centering
    \includegraphics[width=1\columnwidth]{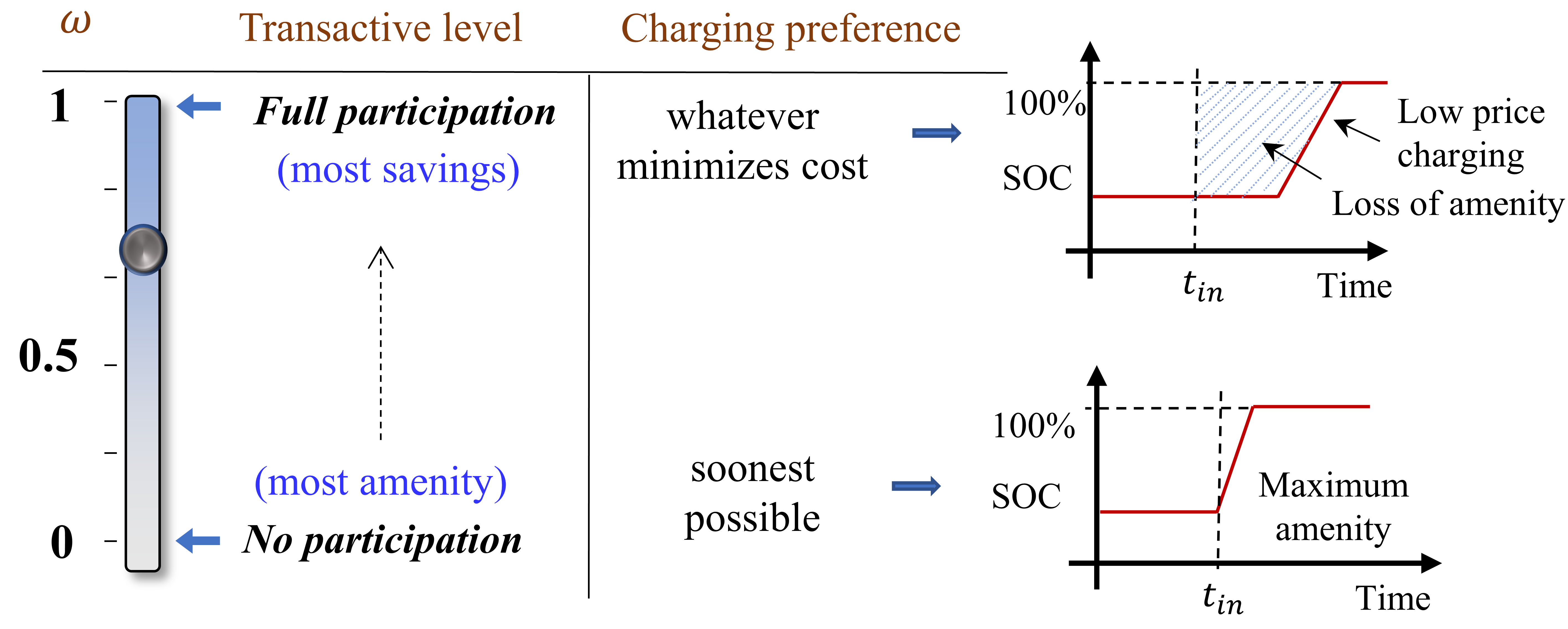}
    \caption{Illustration of how preference slider ($\omega$) acts in 2 extreme cases, i.e., $\omega=0$ (max amenity) and $\omega=1$ (max savings).}
    \label{fig:slider}
    \vspace{-5mm}
\end{figure}

To incorporate a customer's willingness to participate in the transactive market, we introduce the concept of a slider, $\omega$, that each customer can set to their preferred value on the scale of 0 to 1. A higher $\omega$ means the customer is willing to trade their amenity for economic gain in the transactive market. Similarly, a lower $\omega$ reflects that the customer values their amenity more than the cost savings. In the case of an EV agent, we qualitatively define the customer's amenity as  vehicle readiness i.e. availability of a fully charged EV whereas savings come from the reduction in the EV charging cost by deferring charging to the lower price intervals. For the purpose of modeling, we can say that the maximum amenity case (i.e., $\omega=0$) prefers the EV to get charged as soon and as fast as possible after arriving home.
In contrast, the maximum savings case (i.e., $\omega=1$) prefers to minimize the cost while charging as long as the car is fully charged before leaving home.
\figurename \ref{fig:slider} illustrates these two extreme scenarios to explain the impact of $\omega=0$ (max amenity) and $\omega=1$ (max savings). We will develop the mathematical model of this tradeoff in the next section.

\subsection{Objective Function}
\begin{figure}
    \centering
    \includegraphics[width=1\columnwidth]{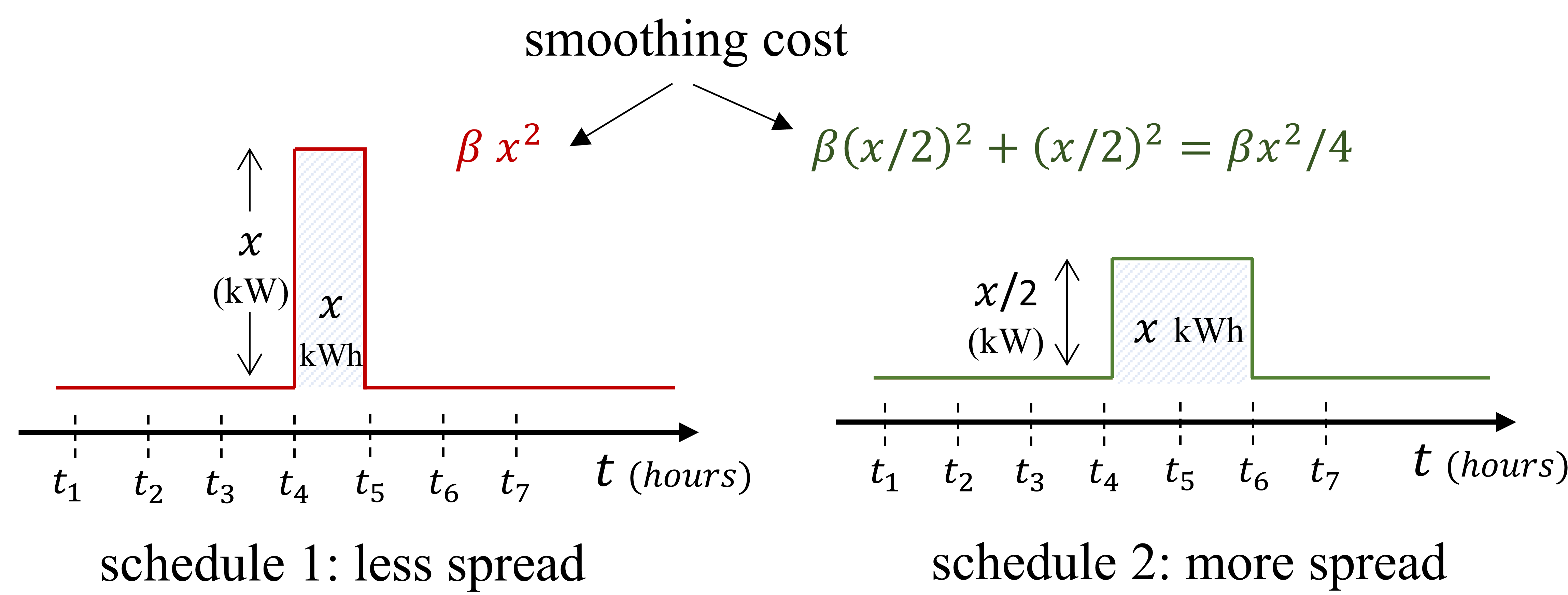}
    \vspace{-9mm}
    \caption{Illustration of the concept of smoothing term in the cost function to encourage charging spread in the optimal schedule}
    \label{fig:beta_concept}
    \vspace{-0mm}
\end{figure}

Mathematically, the net operational cost to the customer (that should be minimized) can be expressed as follows:
\begin{equation}
\begin{gathered}
    Cost = \\
    \omega \sum_{t \in T}  P_f(t)\left(E^{in}_{b}(t)-E^{out}_{b}(t)\right) + \phi \left(E^{out}_{b}(t)+E^{in}_{b}(t)\right)\\
    + (1-\omega) \sum_{t \in T} \alpha \left(C^{max}-C(t)\right)\\
    + \sum_{t \in T} \beta \left(E^{out}_b(t)+E^{in}_b(t)\right)^2
\end{gathered}
\label{eq:obj}
\end{equation}
where $P_f(t)$ denotes the forecast electricity price for hour $t$. In the cost expression, the first term represents the cost incurred due to charging and discharging the vehicle while accounting for the battery degradation cost as well. The EV battery degradation cost is modeled as a constant rate $\phi$ in \$/kWh. The first term in \ref{eq:obj} is multiplied by the slider, $\omega$, to ensure higher transactive preference leads to higher cost savings. The second term in \ref{eq:obj} represents the discomfort cost or the cost assigned to loss of amenity. This term minimizes the amenity loss by encouraging EV to charge fully as fast as possible. The amenity needs to increase when there is a lower preference for cost savings; therefore, it is multiplied by $(1-\omega)$. $\alpha$ is a constant to model the customer's inconvenience cost in \$/kWh. The third term is a with smoothing coefficient $\beta$ encourages the TEV agent to spread its charging or discharging schedule over multiple intervals to avoid volatility. For instance, if an EV needs $x$ kWh energy to fully charge, the smoothing term will prefer a charging schedule with more spread, i.e., $(x/2)^2+(x/2)^2 \le x^2$ as illustrated in \figurename \ref{fig:beta_concept}. This is important because it discourages agents to respond homogeneously to price and thus avoids synchronized behavior such as large spikes or valleys in the net feeder load that can have destabilizing market consequences between market iterations. In practice, the $\beta$ may be one of the regulatory mandates from DSO. As the price forecast is not perfect and is impacted by EV market bidding, an intelligent agent should hedge its bids by spreading them to cover a large range of possibilities. 
\subsection{Day-Ahead Optimization}
\label{Sec:Optimization}
The optimization problem is formulated for each TEV agent separately to get an optimal charging and discharging schedule for a time horizon $T$. Based on the discussion so far, a complete optimization formulation can be written as:
\begin{equation}
\begin{aligned}
    \min_{E^{out},E^{in}} \quad & Cost \\
    \text{subject to} \quad & Eq. (\ref{eq:first_const})-(\ref{eq:last_const})
    \label{eq:opt}
\end{aligned}
\end{equation}
The proposed scheduling can support both EV technologies: V1G and V2G. In V1G, the TEV is restricted to only charge the vehicle from the grid and not allowed to bid the net export of energy from the vehicle to the grid i.e. $E_{max}^{out}$ is set to zero. Whereas in V2G, the TEV can bid both net import and net export of energy. A comparative study of V1G and V2G will also be discussed later in the Section \ref{sec:results}. The obtained optimal schedule is utilized to construct a bid and participate in the market as discussed in the next section.

\begin{figure}
    \centering
    \includegraphics[width=1\columnwidth]{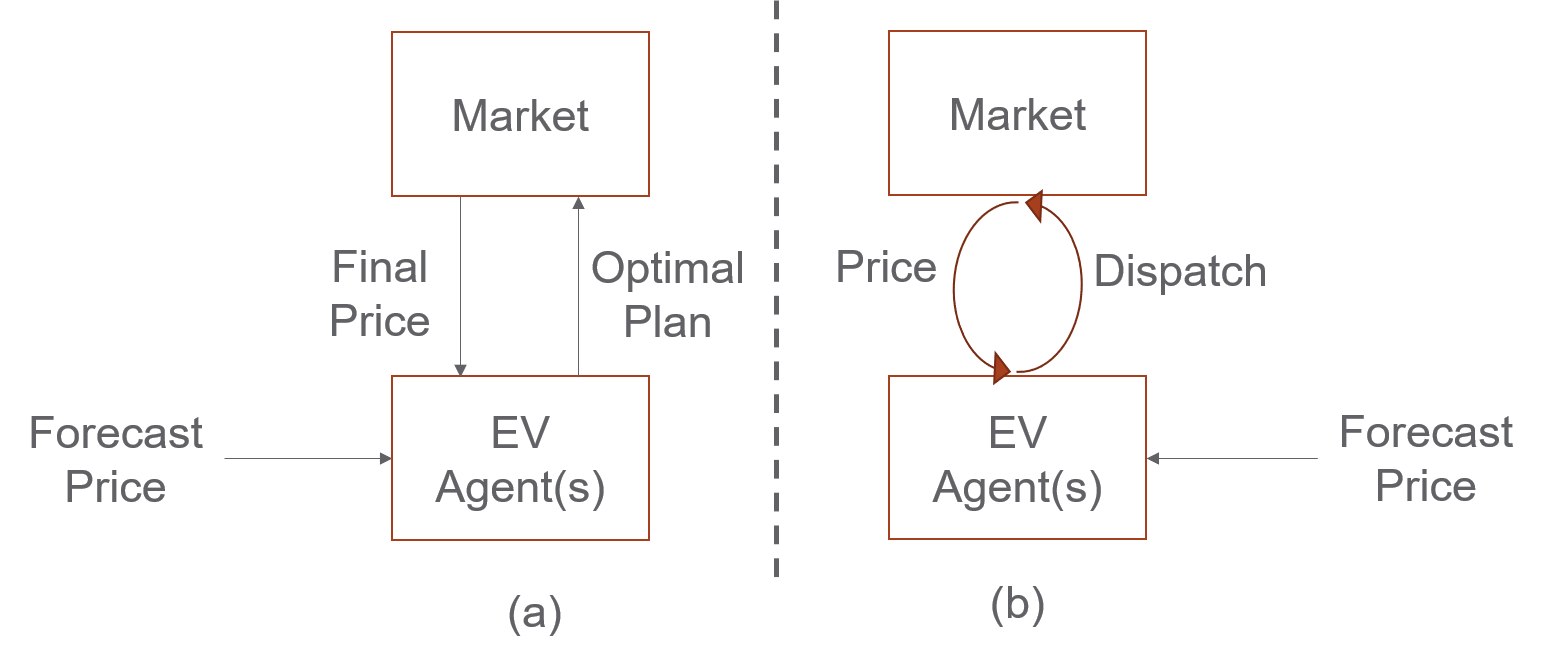}
    \vspace{-15pt}
    \caption{Illustration of EV Agent behavior under (a) price taking setting and (b) price making setting.}
    \label{fig:ev_agent_market_motivation}
    \vspace{-3mm}
\end{figure}
\section{Market Bidding Mechanism and EV Control}
\textsc{Motivation.} Previous sections explained the process of creating an optimal energy plan for a TEV agent considering its preference level to participate based on forecasted prices. However, as discussed in Section I,  the price-taking EVs could jeopardize the grid performance \cite{remco2014}. Therefore, we utilize a transactive market mechanism that incorporates EV agents into the price formation process as follows. Fig.~\ref{fig:ev_agent_market_motivation} shows two scenarios where EV agents may be exposed to the market prices. Fig.~\ref{fig:ev_agent_market_motivation}(a) shows a price taking scenario when the agent forecasts the price and submits the operation plan, which is neither optimal with respect to the cleared price from the market (based on which it pays) nor alleviates congestion issues in the grid. In contrast, Fig.~\ref{fig:ev_agent_market_motivation}(b) shows transactive control through a price formation (making) scenario where the agent iterates with the market to effectively remove issues such as congestion as well as incorporates its willingness to consume energy without sharing any private information. The next section presents the mechanism to achieve such transactive control.

\textsc{Market Abstraction.} Note that one of the main goals of this paper is to show generic participation of EVs in the electricity market. Therefore, we do not deeply examine the issues of 1) aggregation methodologies of EVs to participate in the market and 2) the exact rules of retailers and utilities in influencing the EV agent behavior. This is because the exact strategy of aggregation is a regulator and policy question, which must be discussed once generic aggregation strategies have been shown to be adequate (as demonstrated in this paper). Currently, competitive markets exist at the transmission level (wholesale markets), overseen by independent system operators (ISOs). Because ISOs provide rules and regulations for the demand-side resources to participate in the wholesale market, the exact market mechanisms vary from one ISO to another. Nonetheless, these mechanism are usually built on two main market periods, the day-ahead (DA) market and the real-time (RT) market. In DA, the energy purchase commitments are performed to construct a baseline and then delivered in RT by utilizing opportunities arising in the RT market around the baseline.

Following this generic structure, the TEV agent is designed to participate in the DA as well as RT market. Hence, a local retail market operator (DSO in this work) is envisioned to oversee this participation. The TEV agent takes the optimal hourly schedule as a reference for DA market participation and prepares privacy-preserving price-quantity bid curves around the desired operating point. The bid curve is designed such that any difference between the forecast electricity price (that is used to prepare the operational schedule) and actual electricity price (obtained once the market clears) are taken to properly adjust the operation of the EV in real time. To deal with the price volatility and deviations from the day-ahead plan, the bidding process is iterated every hour in the retail DA market by allowing the EVs to rebid the next 48 hours given the updated forecast of retail day-ahead prices. In between these intervals, the TEV participates in the retail RT market that runs every 5 minutes. Eventually, the final part is to ensure consensus among all TEV agents, i.e., the market has converged to an agreed price and quantity. This is important before the local operation of EV agents needs to be communicated to the whoelsale market's DA and RT periods. In this paper, this is done by checking the evolution of DA prices such that they converge before they need to submit their plans to the wholesale market. The details of each of these stages are provided below.
\vspace{-5mm}
\subsection{Day-ahead Market}
\begin{figure}
    \centering
    \includegraphics[width=0.9\columnwidth]{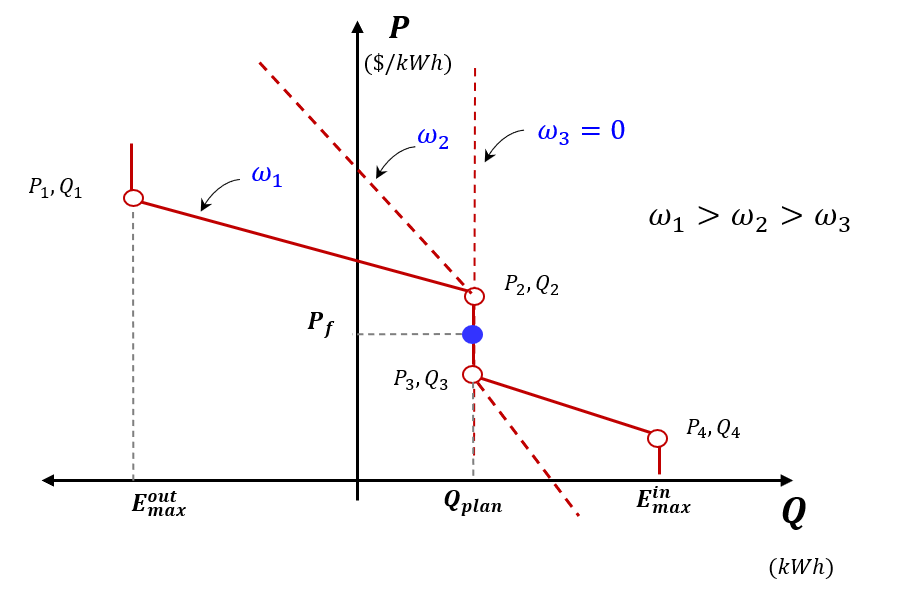}
    \vspace{-10pt}
    \caption{Conceptual illustration of DA four-point bidding for TEV agents with different transactive preferences. Higher slider leads to more flexible bidding curve and vice versa.}
    \label{fig:bid_curve}
    \vspace{-3mm}
\end{figure}

The proposed retail DA market participation involves two successive processes, namely optimal quantity scheduling and bid curve formulation. First, at any given hour, the optimal planned quantity array of $N$ entries, $(Q_{plan}(t)=E^{in}_b(t)-E^{out}_b(t))$, is obtained from the optimization \eqref{eq:opt}, where $t\in T= {1,2,...,N}$. Based on these quantities, $N$ four-point bid curves are constructed, one for each entry $t$ in the DA window, as shown in \figurename \ref{fig:bid_curve}. 
The process of computing all four points  is explained as follows. The slope, $C_{slope}$ and intercept, $C_{intercept}$ of the curve are computed as:
\begin{align}
C_{slope}  &= \left(\frac{\displaystyle\max_{t\in T}(P_f(t)) - \min_{t\in T}(P_f(t))}{-E^{out}_{max} - E^{in}_{max}}\right)\cdot \frac{1}{\omega} \\
C_{intercept}(t)  &= P_f(t) - C_{slope}\cdot Q_{plan}(t)
\end{align}
where, $P_f(.)$ is a forecast price array with $N$ entries. The 4 points on the bid curve, shown in \figurename \ref{fig:bid_curve}, are calculated as:
\begin{align}
   &\begin{cases}
   P_1(t) =&  -E^{out}_{max} \cdot C_{slope}  + C_{intercept} (t) + db \\
Q_1(t) =& -E^{out}_{max}
  \end{cases}\nonumber \\
    & \begin{cases}
   P_2(t) =& Q_{plan}(t) \cdot C_{slope}  + C_{intercept} (t) + db \\
Q_2(t) =& Q_{plan}(t)
  \end{cases}\nonumber\\
   &    \begin{cases}
   P_3(t) =& Q_{plan}(t) \cdot C_{slope}  + C_{intercept} (t) - db \\
Q_3(t) =& Q_{plan}(t)
  \end{cases}\nonumber\\
  &       \begin{cases}
   P_4(t) =& Q_{plan}(t) \cdot C_{slope}  + C_{intercept} (t) - db \\
P4_q(t) =& E^{in}_{max}
  \end{cases}
\end{align}
where $P_i(t)$ and $Q_i(t)$ are the price and quantity, respectively, for $i^{th}$ point on the curve for time $t$. $db$ is the deadband price margin. It is worth emphasizing the role of an individual customer's transactive preference, $\omega$, in bidding as shown in \figurename \ref{fig:bid_curve}. A customer with a lower transactive preference will have a less price-sensitive bid curve, and a non-participating customer ($\omega=0$) will have an inflexible vertical curve, always sticking to the planned quantity. In contrast, a highly transactive customer will adjust its quantity bid in both directions to achieve savings in case price deviates from the forecast.
\subsection{Real-time Market}
Due to the intrinsic uncertainties existing in the system, RT operation deviates from DA. Therefore, the retail RT market performs the adjustments for the deviations and submit its bid every 5 minutes. The RT operation updates its knowledge of the EV SOC from the vehicle every 5 minutes, thus updating the remaining operational flexibility of the EV for RT adjustment.
The day-ahead planned quantity and price obtained from scheduling are directly used for the real-time bid curve formulation. However, the $Q_{plan}$ for real-time bid is updated every 5 minutes by interpolating between the cleared day-ahead quantity of the existing hour and $Q_plan$ for the next hour DA market. Thus, a RT bid curve with the same slope but interpolated quantity is used every 5 minutes for an hour. As the new bids for each current DA period are made, the same procedure repeats.
\vspace{-2mm}
\subsection{Real-time Control}
The control implementation module of TEV takes the clearing price from the retail RT market ($P_{cl}^{RT}$) and controls the EV equipment such that operation meets the committed quantity ($Q^{RT}$) from the RT bid curve submitted to the market. An actuation signal is sent to the EV charger to guide its operation. 
\vspace{-2mm}
\subsection{Market Reconciliation}
The final step in the market component is the market reconciliation. The DA market discussed above has particular times when it accepts bids. For example, a DA market runs every day at 10 AM for products to be committed from next day midnight onward \cite{PJM-day}. From this perspective, it is important to analyze the dynamics of the TEV agent when they iterate through time and whether they diverge from their plans. This can be analyzed by the entity purchasing energy from the wholesale on behalf of EV agents\footnote{As mentioned earlier, description of the exact architecture of aggregation, and the entity governing this exchange is out of scope of this paper.} as follows:
\begin{enumerate}
    \item Forecast hourly price of the energy and the quantity to be purchased from the wholesale market based on historical data along with fixed loads in the distribution grid.
    \item Aggregate bid curves from individual EV agents with inflexible load addition to construct a demand curve for DA market (hourly resolution). 
    \item Clear the retail market by finding an intersection between the forecasted energy price/quantity curve with the aggregated demand bid of TEV agents. If the quantity exceeds the feeder limits, a congestion surcharge\footnote{We do not discuss rate design in this paper, though it will be needed to have equitable final billing for the consumer \cite{rg_pratt_dsot_2021}} can be imposed on the cleared price.
    \item Use the cleared price of the aggregated curve and its mapped quantity for each TEV agent as the base point, and use RT bidding to provide correction opportunities. 
    \item Repeat for each hour.
\end{enumerate}

In the above method, it can be observed that the DSO is iterating with EVs to find a consensus in their charge/discharge plan. In doing so, the EVs, with the help of the DSO, converge to a system-wide agreeable price and quantity. The analysis can be done to observe evolution of prices and whether they converge to the information submitted to the wholesale market. In this way, reconciliation of the proposed TEV agents can be done with the power grid markets.  
\section{Case Study}
\label{sec:results}
\begin{figure}
    \centering
    \includegraphics[width=1\columnwidth]{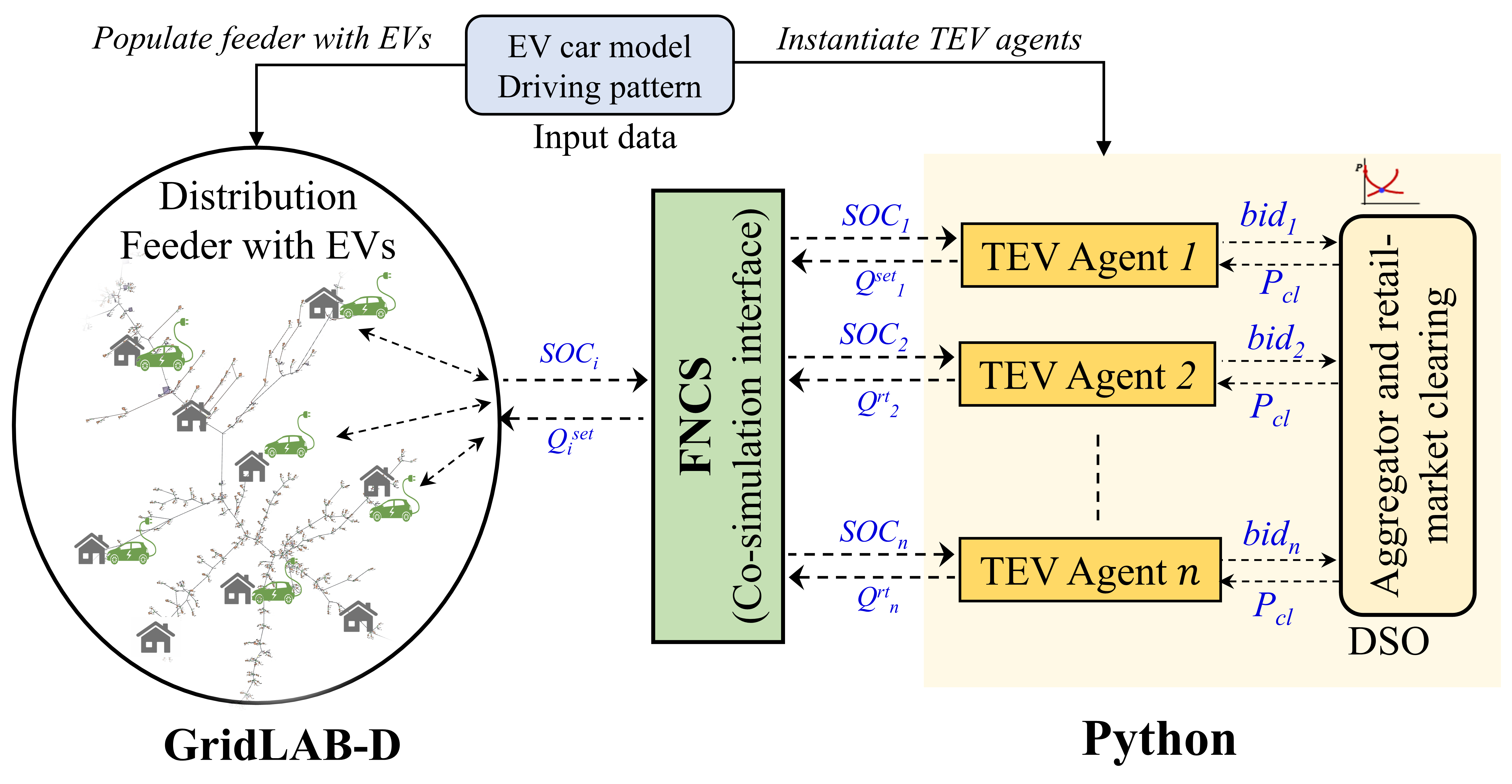}
    \vspace{-20pt}
    \caption{Schematic of the testbed to demonstrate TEV agent performance}
    \label{fig:testbed}
    \vspace{-3mm}
\end{figure}

A co-simulation test-bed was created to demonstrate the TEV agent performance as shown in the \figurename \ref{fig:testbed}. A taxonomy feeder R4-12.47-1 is utilized that represents a heavily populated urban area in a central part of the USA with the primary feeder extending
into a lightly populated rural area \cite{taxonomy_report}. The feeder is populated with around 530 residential houses, out of which 160 randomly selected houses own EVs. The distribution feeder and houses with EVs are modeled in detail using GridLAB-D -- an open source power flow tool. TEV agents are modeled in python, and each of them communicates to their corresponding EV (sends charging set-point and receives SOC) in GridLAB-D via FNCS, a co-simulation platform designed to integrate and coordinate various power system simulation tools. The DSO aggregator and retail market are also modeled in python. The physical parameters for various EV models are provided in the Appendix. A random slider, $\omega$, between 0 to 1 is assigned to each EV. 

The proposed TEV performance is compared with the base case when EVs do not participate in a transactive market and start charging as soon as they arrive home. 
To quantify the performance, we define two indices, 'Savings' and 'Amenity' for each TEV agent w.r.t. the base case over the entire simulation time range as follows.
\begin{gather*}
Savings = \% \text{ reduction in the EV charging bill w.r.t. base case}\\ 
Amenity = \% \text{ of hours the EV is fully charged w.r.t. base case}
\end{gather*}
A 100\% $Amenity$ means that the availability of a fully charged EV is same as the base case, and there is no loss of amenity. In total 30 days are simulated out of which first 2 days are discarded for warming up the forecast.
The following subsections discusses the various aspects of TEV and market performance.


\subsection{Market performance}

\begin{figure}
    \centering
    \includegraphics[width=1\columnwidth]{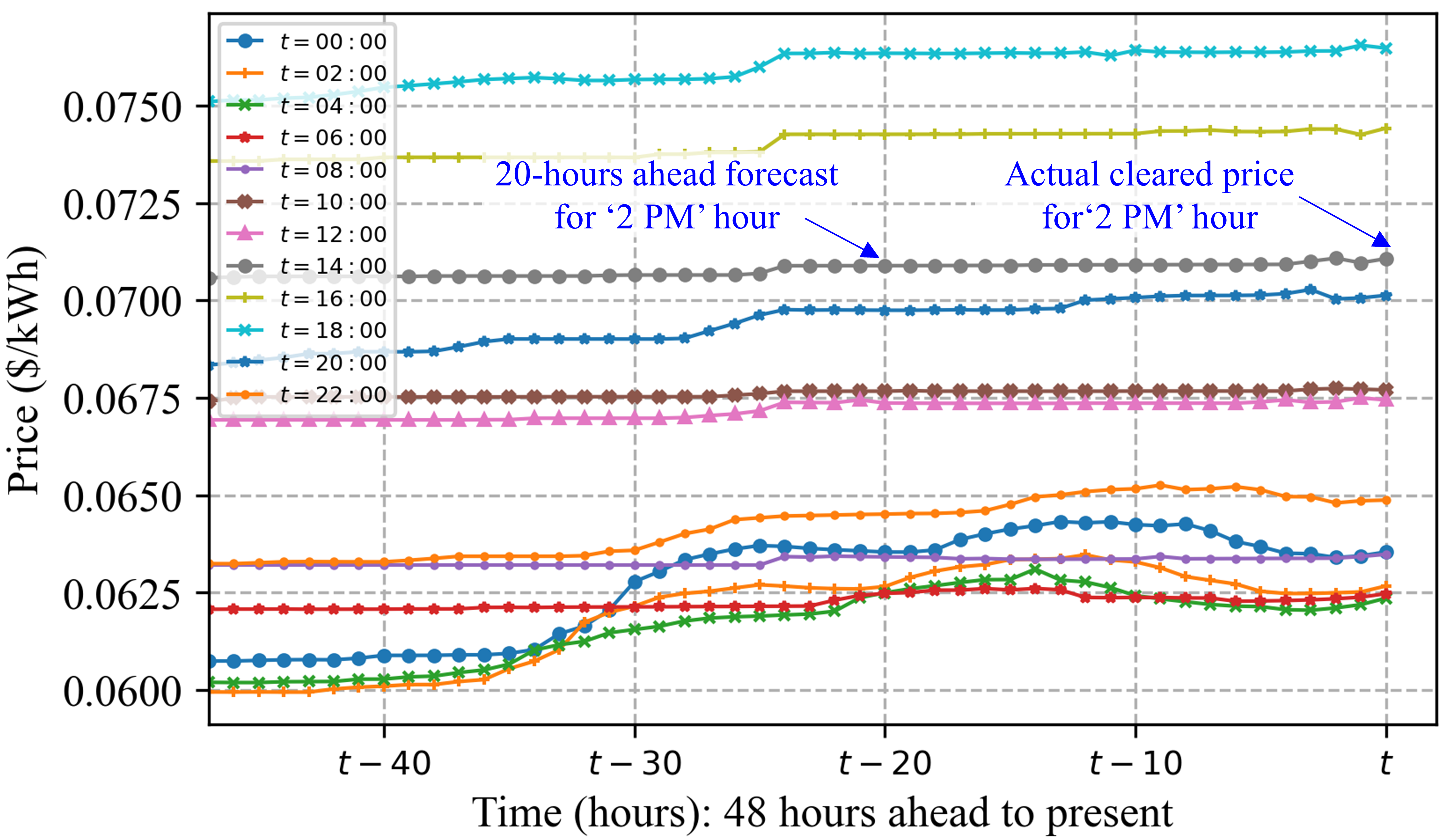}
    \vspace{-20pt}
    \caption{Evolution of DA cleared prices for each hour over the DA horizon window of 2 days}
    \label{fig:convergence}
    \vspace{-3mm}
\end{figure}

\figurename~\ref{fig:convergence} shows the evolution of the cleared prices of different hours denoted by various $t$ values for $4^{th}$ day as the TEV agents iterate their price-bids in the DA market throughout the horizon window of 48 hours i.e. ($t-47$) to $t$ (actual price). Price corresponding to ($t-i$) denotes the cleared price for $t^{th}$ hour at $i$ hours advance. It can be seen that prices stabilize within the horizon window, verifying the initialization of the market 48 hours before the actual submission of the bid is sufficient for consensus. It is worth noting in \figurename \ref{fig:convergence} that the prices for night hours (12 midnight -- 4 AM) take relatively longer to stabilize due to those being low-price transactive hours for most TEV agents.
This shows the importance of market reconciliation process.

\begin{figure}
    \centering
    \includegraphics[width=1.0\columnwidth]{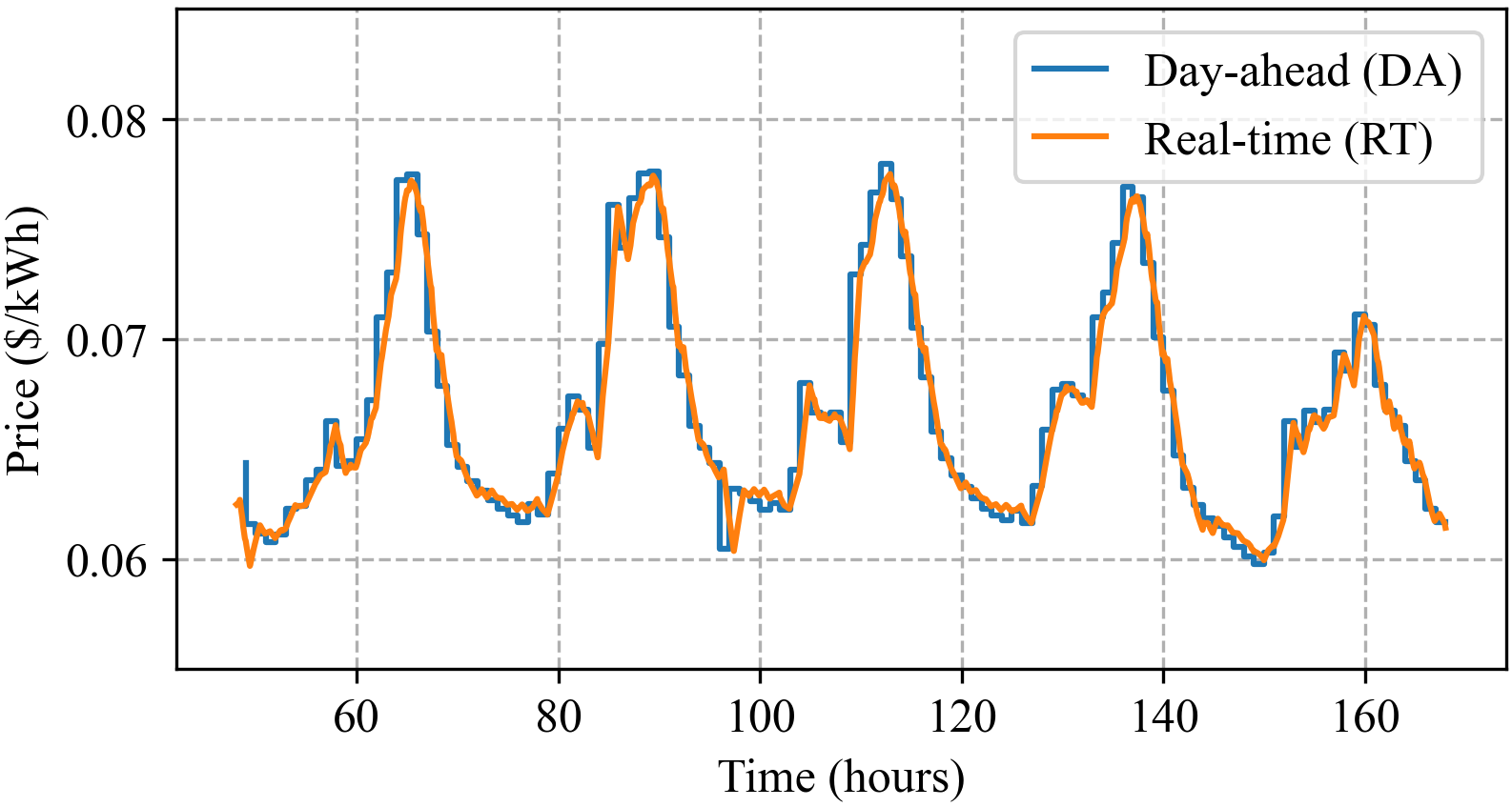}
    \vspace{-20pt}
    \caption{Real-time (RT) cleared prices closely follow the day-ahead (DA) market cleared prices with TEV agents participation.}
    \label{fig:da_rt_price}
    \vspace{-3mm}
\end{figure}

In \figurename~\ref{fig:da_rt_price}, real-time uncertainties and grid constraints have been removed to show market behavior of both day-ahead and real-time markets. From \figurename~\ref{fig:da_rt_price}, it can be seen that the cleared day-ahead and real-time market prices are very close to each other. This verifies the design of DA and RT markets, where in the absence of any unforeseen circumstances, both markets produce similar prices.

\color{black}

\subsection{Impact of Slider Preference on EV Customer}
\begin{figure}
    \centering
    \includegraphics[width=1.0\columnwidth]{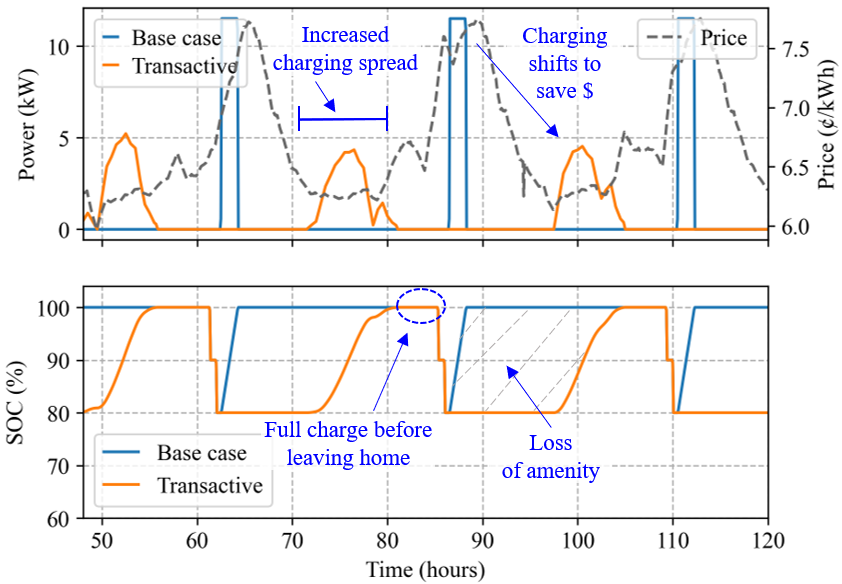}
    \vspace{-20pt}
    \caption{Charging and SOC profile of a TEV with high slider setting and its comparison with base case}
    \label{fig:high_slider_ev}
    \vspace{-0mm}
\end{figure}

\begin{figure}
    \centering
    \vspace{-4mm}
    \includegraphics[width=1\columnwidth]{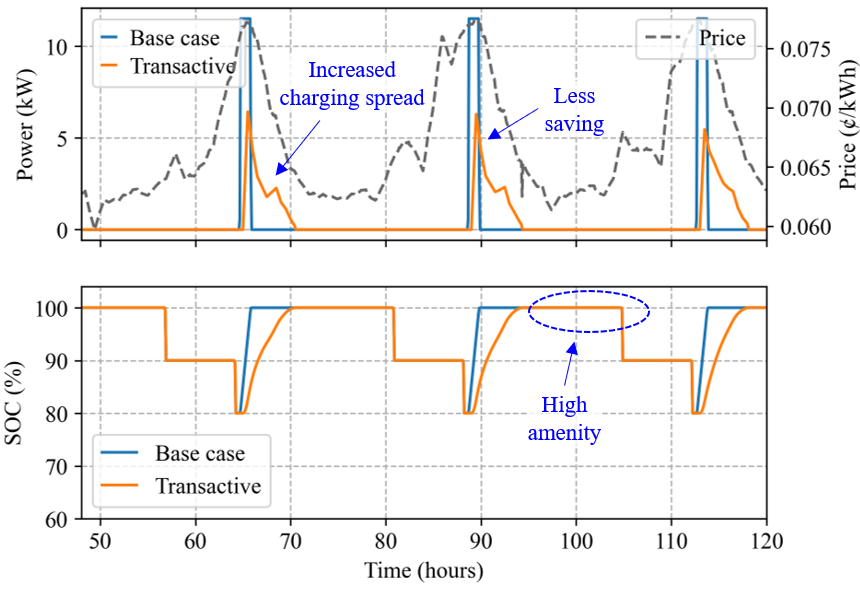}
    \vspace{-20pt}
     \caption{Charging and SOC profile of a TEV with low slider setting and its comparison with base case}
    \label{fig:low_slider_ev}
    \vspace{-6mm}
\end{figure}

The slider setting reflects the transactive preference of an EV owner. For clarity, two extreme behaviors with very high and very low sliders are shown here. The charging behavior of a TEV with high slider setting, $\omega=0.9$, is compared with base case in \figurename \ref{fig:high_slider_ev}. It can be seen in \figurename \ref{fig:high_slider_ev} (top) that TEV is able to shift the charging to lower price duration (midnight) to achieve savings. Nonetheless, it also leads to reduced duration with full charge compared to base case reflected in lower amenity value as shown in \figurename \ref{fig:high_slider_ev} (bottom). Note that the EV gets fully charged before the time it leaves home, and the charging is more spread out compared to the base case due to effect of smoothing coefficient, $\beta$. In contrast, the charging behavior of an EV with lower slider, $\omega=0.2$, is shown in \figurename \ref{fig:low_slider_ev}, where the EV prefers amenity over savings and does not fully shift its charging to lower price intervals. This results in lesser savings and higher amenity.

\begin{figure}
    \centering
    \includegraphics[width=1\columnwidth]{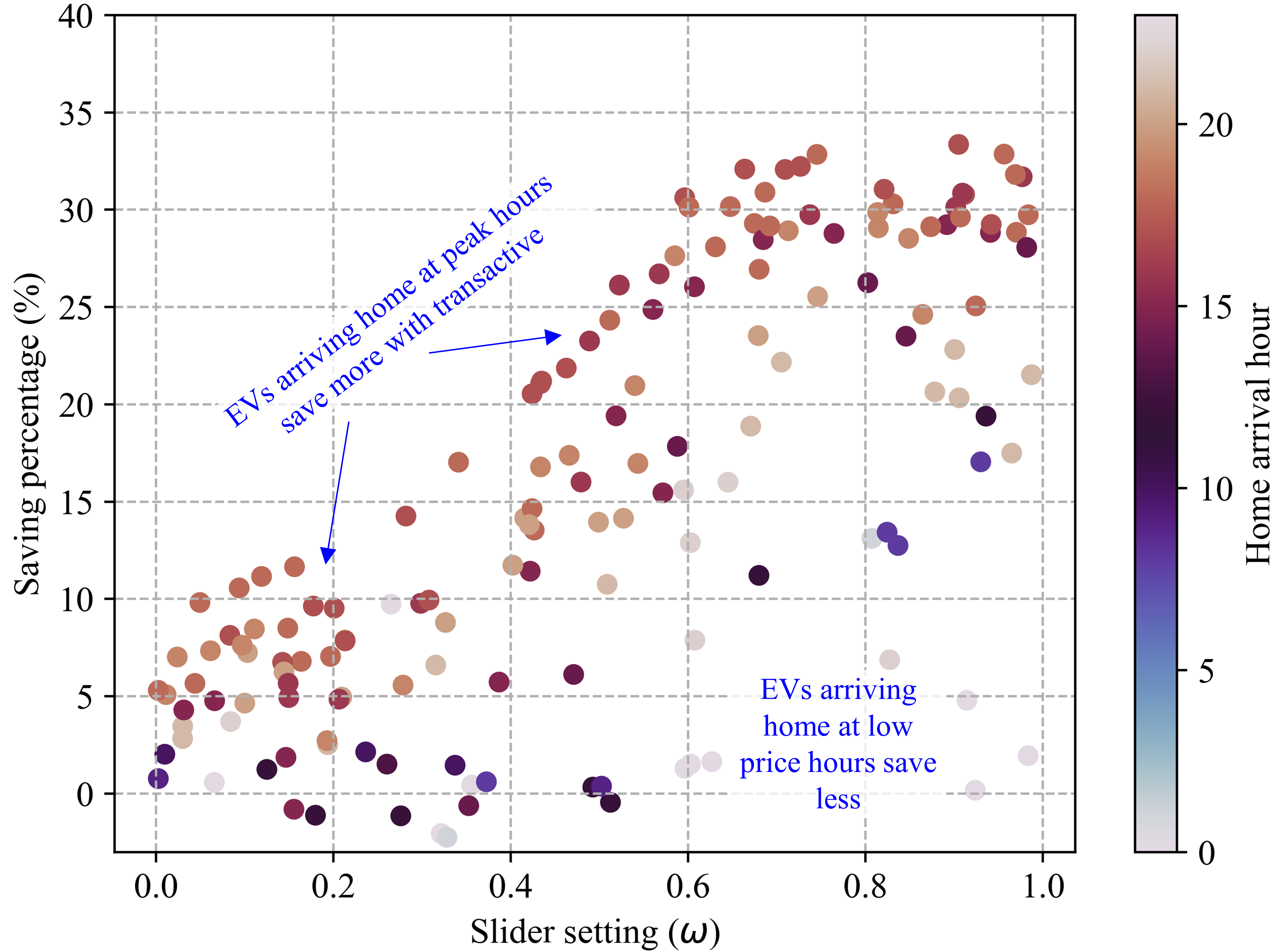}
    \vspace{-20pt}
    \caption{Distribution of savings incurred by all TEVs with respect to their slider settings}
    \label{fig:saving_slider_dist}
    \vspace{-0mm}
\end{figure}

\begin{figure}
    \centering
    \includegraphics[width=1\columnwidth]{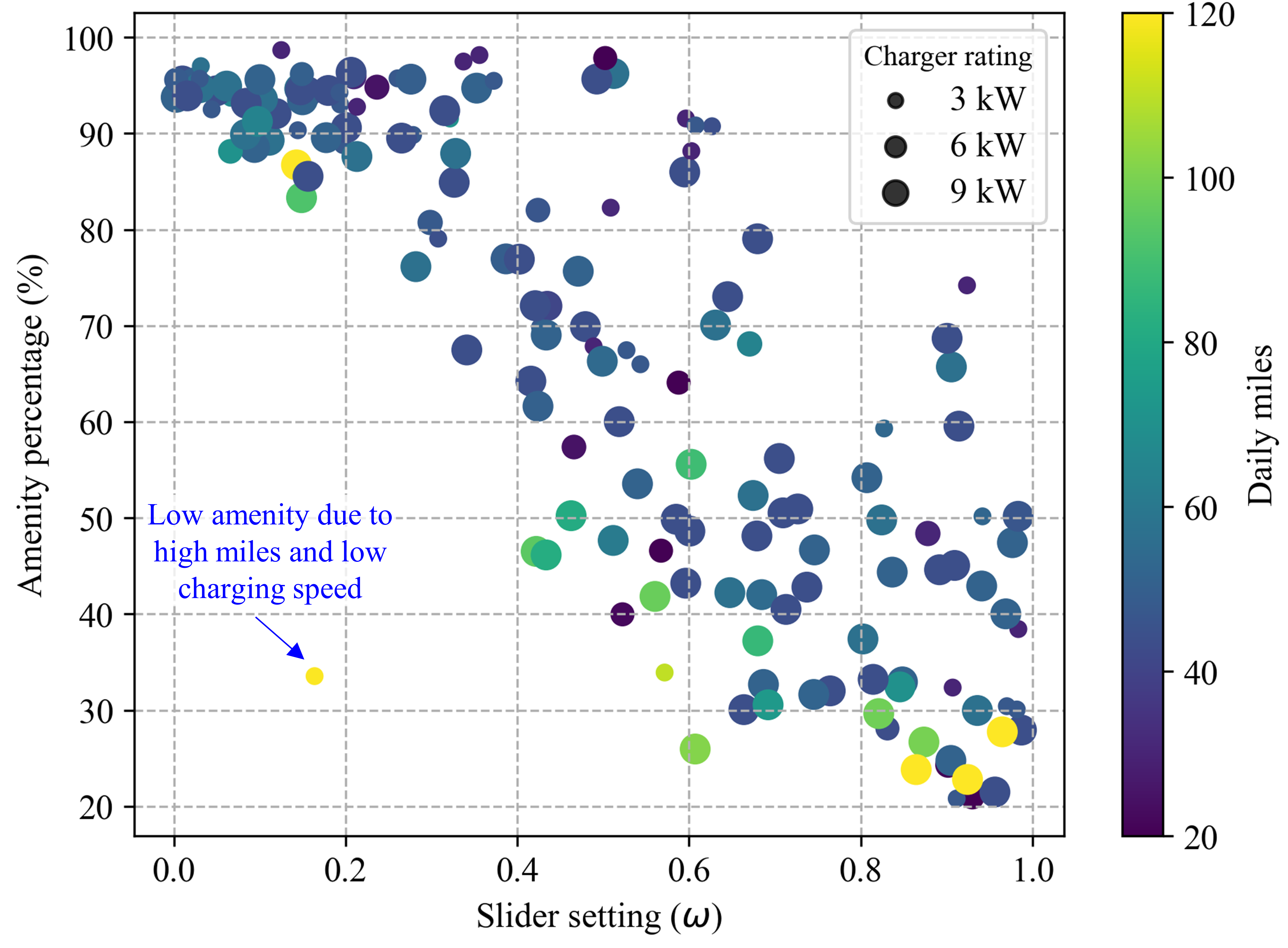}
    \vspace{-20pt}
    \caption{Distribution of amenity provided by all TEVs with respect to their slider settings}
    \label{fig:amenity_slider_dist}
    \vspace{-3mm}
\end{figure}

\figurename~\ref{fig:saving_slider_dist} shows the distribution of $Savings$ of all TEVs with respect to their slider setting. An increasing trend of savings with higher slider can be observed here. It should be noted that an EV's driving pattern and charger specifications also affect it's savings. For instance, EVs' home arrival time is one such major factor identified on \figurename \ref{fig:saving_slider_dist} with different colors. EVs arriving home at the peak price hours (evening hours) are able to save more by shifting their charging to the lower price duration compared to the EVs who were already arriving and charging at low price hours in day time. Nonetheless, the average savings of EV fleet improves with transactive participation. Similarly, \figurename \ref{fig:amenity_slider_dist} shows the distribution of $Amenity$ of all EVs with respect to their slider choice. A decreasing amenity trend with higher $\omega$ can be observed where an EV with slider close to 0 has almost 100\% amenity, indicating no loss of amenity compared to base case. Just like savings, amenity also depends on several other factors. \figurename~\ref{fig:amenity_slider_dist} shows the influence of two such factors (i.e. charger rating and daily travel miles of the EV) via the circle size and color. EVs with high travel miles (shown by yellow-green color) and slower charger (smaller circles) will take relatively longer to fully charge compared to their peers with similar $\omega$, and thus may have lower amenity. 

Overall, the proposed TEV agent is able to honor the customer's transactive preference by trading off savings and amenity appropriately.




\subsection{Impact of TEV at the System Level}
\begin{figure}
    \centering
     \vspace{-10pt}
    \includegraphics[width=0.9\columnwidth]{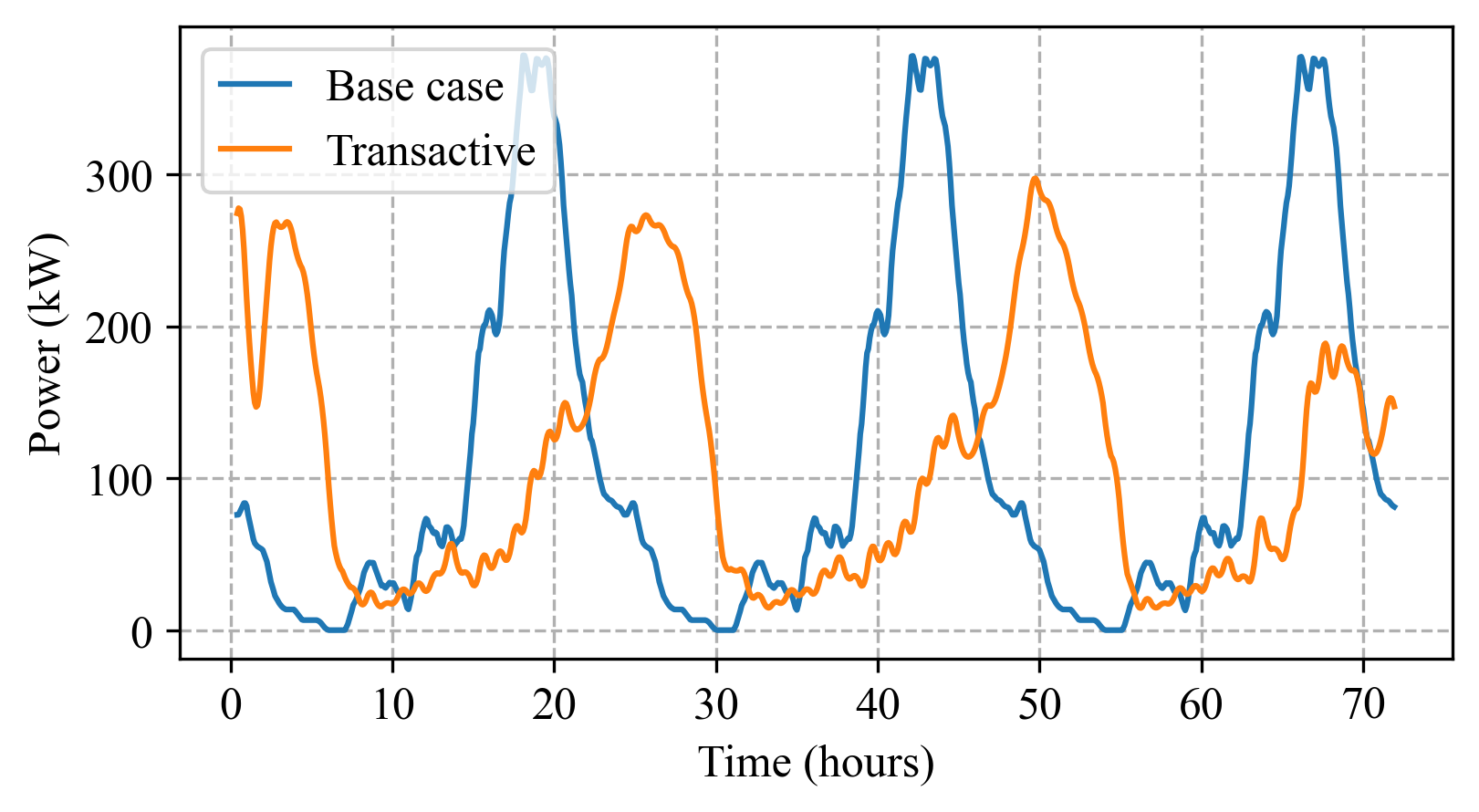}
    \includegraphics[width=0.9\columnwidth]{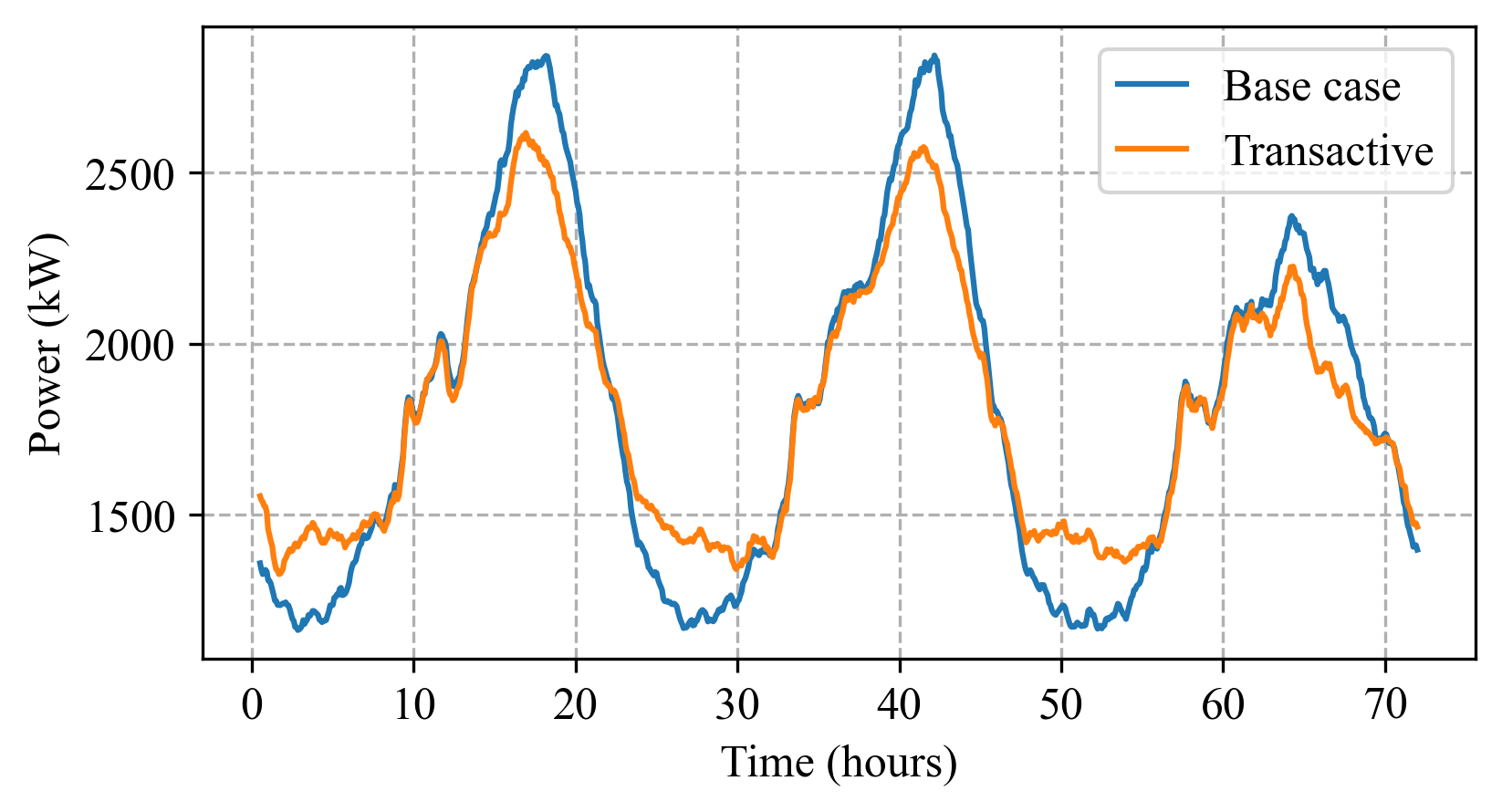}
    \vspace{-10pt}
    \caption{Impact of transactive market on aggregated EV charging profile (top) and total substation load profile (bottom)}
    \label{fig:utility_v1g}
    \vspace{-0mm}
\end{figure}

\begin{figure}
    \centering
    \includegraphics[width=0.9\columnwidth]{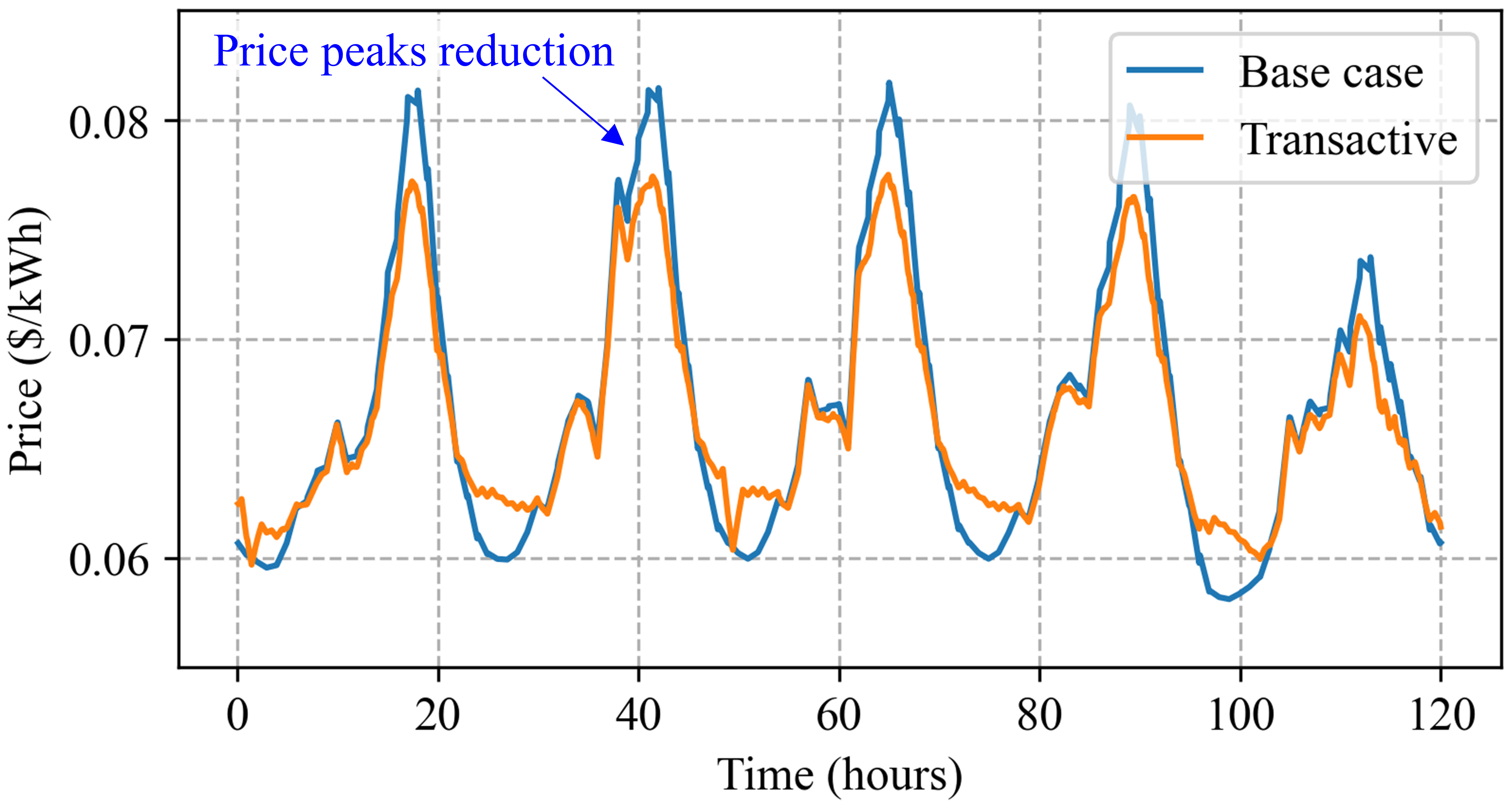}
    \vspace{-10pt}
    \caption{Impact of TEV agents participation on the RT market clearing prices compared to their non-transactive base cases}
    \label{fig:price_compare}
    \vspace{-3mm}
\end{figure}

The aggregate charging behavior of all TEVs is compared with the base case in \figurename~\ref{fig:utility_v1g} (top) where the charging peak is reduced as well as shifted from peak load hours to midnight. This results in the peak load reduction and valley filling of the net substation load profile, leading to load flattening as shown in \figurename \ref{fig:utility_v1g}(bottom). Furthermore, the TEV agents are also able to reduce the RT cleared price peaks as shown in \figurename~\ref{fig:price_compare} due to their \textit{'price making'} formulation. The reduction of prices and peak load are beneficial for both the customers as well as the distribution utility.



\subsection{Impact of Smoothing Coefficient}
\begin{figure}
    \centering
    \vspace{-10pt}
    \includegraphics[width=0.95\columnwidth]{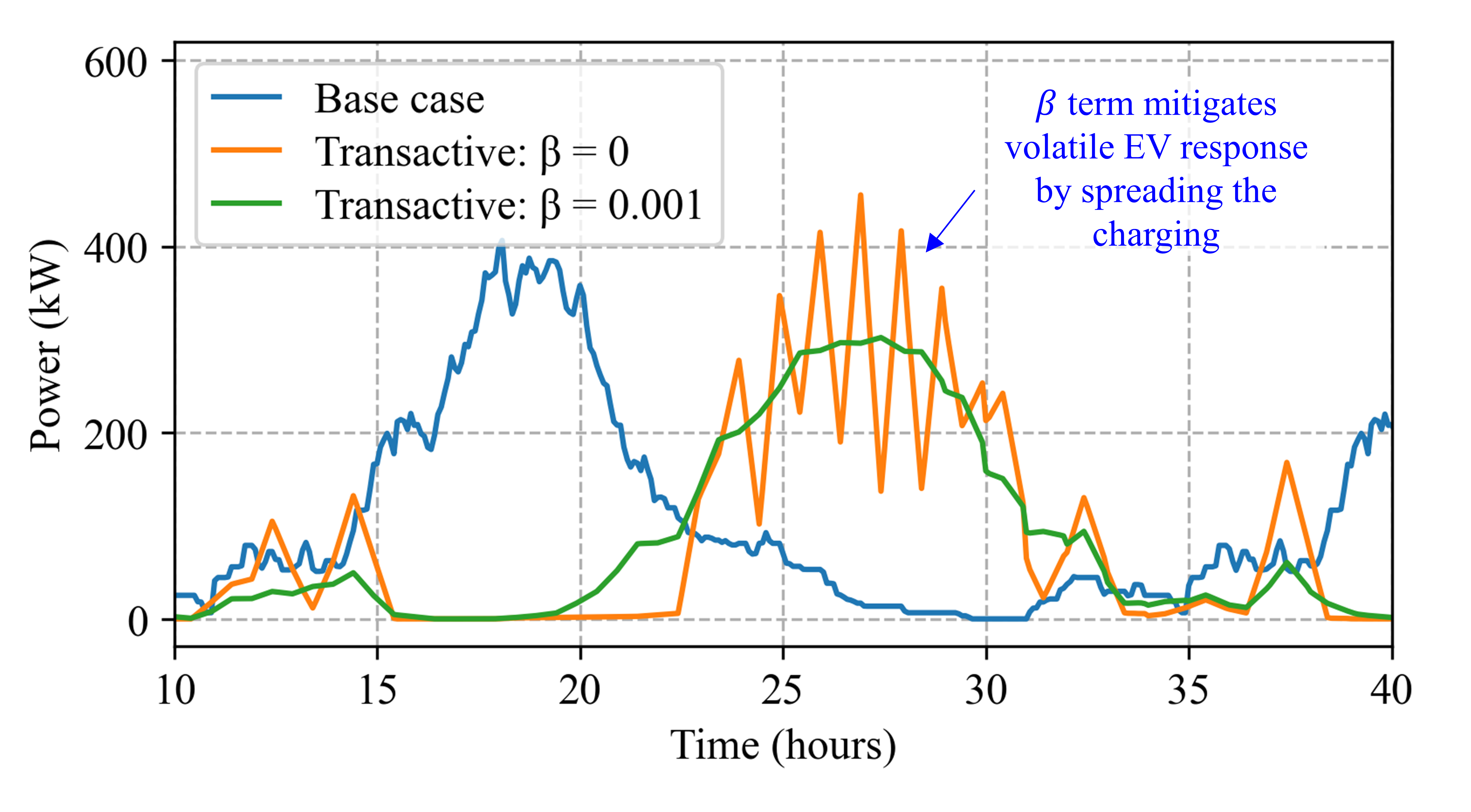}
    \vspace{-15pt}
    \caption{The impact of smoothing term $\beta$ in mitigating the synchronous response of all TEVs}
    \label{fig:beta_ev_agg}
    \vspace{-3mm}
\end{figure}

In order to show the true value of including $\beta$ in TEV optimization, we simulated a case where all EVs are highly transactive, i.e., with $\omega \geq 0.8$. It can be seen in \figurename \ref{fig:beta_ev_agg} that with $\beta=0$, all TEVs synchronize their response to maximize savings at the same time, leading to a highly volatile and undesirable aggregate EV profile. Whereas with $\beta=0.001$, all TEVs spread their charging to mitigate the synchronization issue in net load profile.

\subsection{Comparative benefits of V2G}
\begin{figure}
    \centering
    \includegraphics[width=1\columnwidth]{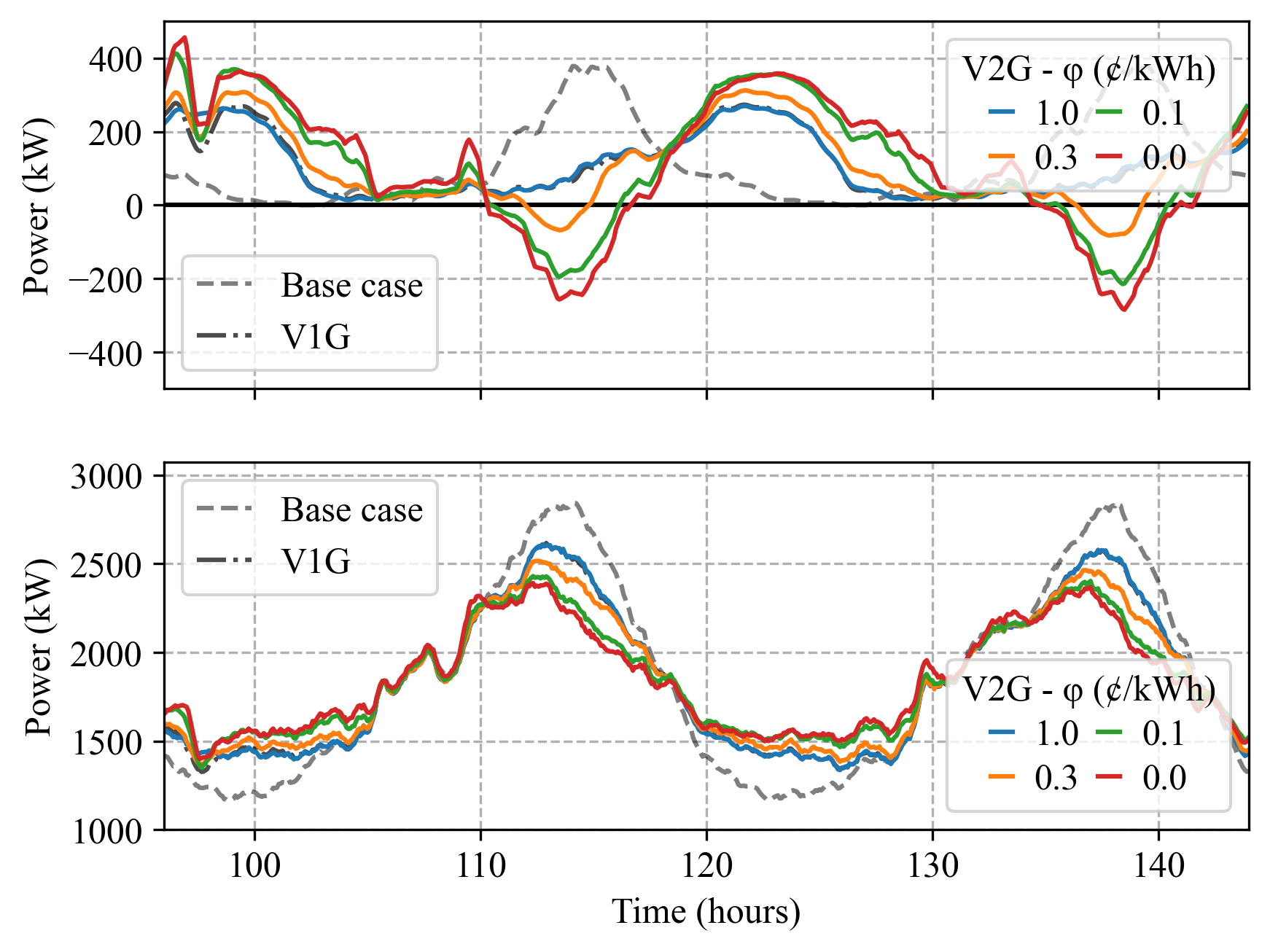}
    \vspace{-20pt}
    \caption{Comparison of aggregate TEV performance (top) and substation load impact (bottom) with V2G technology under different battery degradation cost assumptions.}
    \label{fig:v2g_power}
    \vspace{-1mm}
\end{figure}

\begin{figure}
    \centering
    \includegraphics[width=0.9\columnwidth]{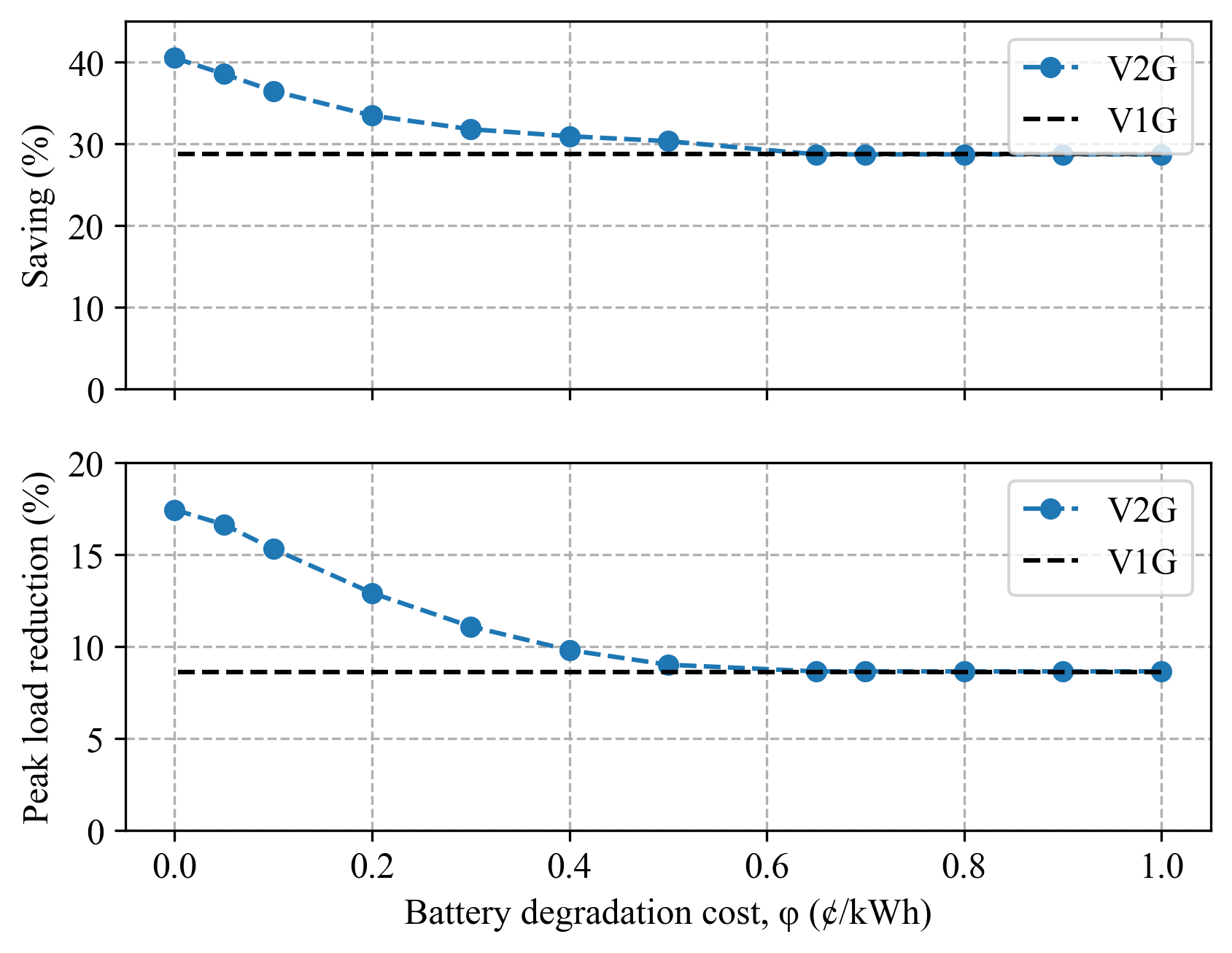}
    \vspace{-12pt}
    \caption{Comparative benefits of V2G with respect to V1G in customer savings and utility peak load reduction}
    \label{fig:v2g_benefit}
    \vspace{-3mm}
\end{figure}

Because most EVs in present times have significantly higher capacity than required for daily travel, most of the SOC remains underutilized. The remaining SOC can be utilized by V2G mode to increase customer savings as well as achieve load flattening at the utility level. However, the V2G mode reduces the battery life due to increased charging/discharging cycles, which is incorporated in TEV optimization via battery degradation cost, $\phi$. \figurename~\ref{fig:v2g_power} (top) and (bottom) show the aggregated EV charging behavior and corresponding impact on the substation load, respectively, for different values of $\phi$. It can be observed that with lower $\phi$, the V2G mode is more effective by discharging a higher amount in high price hours leading to both higher saving and higher peak load reduction. \figurename~\ref{fig:v2g_benefit}(top) and (bottom) compare the average savings and total peak load reduction of V2G mode with V1G mode for different degradation cost values. It can be observed that V2G starts providing additional benefits on top of V1G only if $\phi$ is less than 0.5 cents/kWh. Currently, the $\phi$ turns out to be in range of 0.8--1.5 cents/kWh, but it is expected to decrease with increases in battery lifetime and  decreases in battery cost. This indicates that the V2G mode has high potential for providing additional benefits to both customers and the utilities in the future as  battery degradation costs decline.




\section{Conclusion}
\label{sec:conclusion}
In this work, a complete design of a TEV agent framework was proposed to provide grid-friendly as well as customer-friendly EV charging scheduling with the following novel features: (a) characterize EV owner's amenity as vehicle readiness, (b) represent customer's willingness to participate in the transactive market via a preference slider, i.e., a preferred trade-off between amenity and savings. A privacy-preserving bidding process that incorporates the customer's transactive preference level is discussed with a detailed transactive market mechanism that integrated the TEV into retail market and reconciles with the current DA and RT structure. Finally, a distribution feeder was utilized as a case study to successfully demonstrate the TEV agent performance on a diverse EV population. It was verified that the customers who chose higher and lower slider values were able to achieve higher savings and higher amenity, respectively while reducing the feeder peak demand. The markets are shown to be converged and have positive impacts on peak pricing and energy consumption. Furthermore, a comparative study on V1G and V2G technologies revealed that the additional benefit of V2G technology over V1G is sensitive to EV battery degradation costs. With current costs, V2G offers relatively less additional benefit but can be promising in the future with a reduction in battery costs.
\vspace{-1mm}

\bibliography{zotero_ankit_ref, citation}

\bibliographystyle{IEEEtran}
\balance
\end{document}